\documentclass[aps,twocolumn,showpacs]{revtex4}
\usepackage{epsf}
\begin{document}
\title{Shattering Transitions in Collision-Induced Fragmentation}
\author{P.~L.~Krapivsky} \email{paulk@bu.edu}
\affiliation{Center for Polymer Studies and Department of Physics,
Boston University, Boston, MA 02215}
\author{E.~Ben-Naim} \email{ebn@lanl.gov}
\affiliation{Theoretical Division and Center for Nonlinear Studies,
Los Alamos National Laboratory, Los Alamos, NM 87545}
\begin{abstract}
  We investigate the kinetics of nonlinear collision-induced
  fragmentation.  We obtain the fragment mass distribution
  analytically by utilizing its travelling wave behavior.  The system
  undergoes a shattering transition in which a finite fraction of the
  mass is lost to infinitesimal fragments (dust).  The nature of the
  shattering transition depends on the fragmentation process.  When
  the larger of the two colliding fragments splits, the transition is
  discontinuous and the entire mass is transformed into dust at the
  transition point. When the smaller fragment splits, the transition
  is continuous with the dust gaining mass steadily on the account of
  the fragments. At the transition point, the fragment mass
  distribution diverges algebraically for small masses, $c(m)\sim
  m^{-\alpha}$, with $\alpha=1.20191\ldots$
\end{abstract}
\maketitle

\section{Introduction}

Fragmentation occurs in numerous physical phenomena and industrial
processes \cite{har,shr,hab,red,ms}. Examples include breakup of
liquid droplets \cite{liq} and atomic nuclei \cite{nuc}, polymer
degradation \cite{ziff}, shattering of solid objects \cite{im,odb},
meteor impacts, and mineral grinding.  Idealized models of such
physical phenomena are also useful conceptual tools for describing
complex systems such as fluid turbulence, spin glasses \cite{sg},
genetic populations \cite{h,djm}, and random Boolean networks
\cite{kauf,flyv}.

In some cases, for example in polymer degradation, the evolution of a
fragment depends only on its size. Therefore, fragments do not
interact and such processes are inherently {\em linear}.  In other
cases including grinding processes, explosions in an enclosed volume,
and breakup of eddies in a turbulent flow \cite{pietro}, interactions
between fragments are essential.  Such fragmentation processes are
intrinsically {\em nonlinear} \cite{sri,sid,kk,laur}. In this study,
we show that the nature of the mass distribution changes qualitatively
due to nonlinearities.

We investigate a basic class of nonlinear fragmentation processes
where binary collisions are the cause of breakage. We show that such
processes exhibit a shattering transition where infinitesimal
fragments (dust) carry a finite fraction of the mass in the system.
We consider the simplest realization where one of the two colliding
fragments breaks into two pieces.  Generically, the number of
fragments diverges in a finite time, indicating shattering into dust.

The nature of the shattering transition depends sensitively upon the
details of the fragmentation process, in particular, which of the two
colliding particles splits. We investigate three possibilities: (A)
either, (B) the larger, and (C) the smaller of the two particles
breaks into two fragments upon collision. In the first two models, as
the transition occurs the entire mass is instantly transformed into
dust. In the third model, the dust mass gradually increases once the
shattering transition occurred.

In contrast with linear fragmentation processes, explicit solutions of
the nonlinear and non-local rate equations are generally not
possible. Nevertheless, the most important physical characteristics
can still be obtained analytically. Interestingly, the fragment mass
distribution attains a travelling wave form as the transition is
approached. Of the spectrum of possible propagation velocities, the
extremal one is selected and it characterizes typical and extremal
behaviors of the mass distribution.  In the case of model C, at the
shattering transition, the mass distribution is algebraic for small
masses, with a transcendental exponent. Past the transition, the
fragment mass distribution approaches a universal form.

We first consider the number density that manifests the shattering
transition (section II). Then, we analyze the fragment mass
distribution using rate equations for a deterministic version (section
III) and a stochastic version (section IV) of the fragmentation
process.  Finally, we summarize our results and outline a few
suggestions for future work (section V).

\section{The number density}

Consider a fragmentation process where at each (binary) collision event, one
particle splits into two pieces while the second particle remains intact.
Without loss of generality, the collision rate is set to unity. Analogous to
the kinetic theory description of collisions in molecular gases, we assume
perfect mixing, namely, absence of spatial correlations between fragments.
The total fragment density, $N(t)$, evolves according to the rate equation
\begin{equation}
\label{N-eq}
\frac{d}{dt}\,N(t)=N^2(t).
\end{equation}
Without loss of generality, the initial density is set to unity,
$N(0)=1$, and therefore, the total density is
\begin{equation}
\label{N-sol}
N(t)=\frac{1}{1-t}.
\end{equation}
In a finite time, the number of fragments diverges and the average
fragment mass vanishes. This divergence indicates that the system
undergoes a shattering transition at $t_c=1$.

Let $\tau=\int_0^t dt'\,N(t')$ be the average number of collisions
experienced by a fragment up to time $t$. This quantity
diverges logarithmically
\begin{equation}
\label{tau}
\tau=\ln N(t)=\ln \frac{1}{1-t}.
\end{equation}
This ``collision counter'' provides a convenient alternative measure
of time.

\section{Deterministic Fragmentation}

To complete the model definition we have to specify which of the
fragments splits, and how it splits. Following Cheng and Redner
\cite{sid}, we consider three possibilities: (A) a randomly chosen,
(B) the larger, and (C) the smaller fragment splits upon collision.
In this section, we consider a deterministic rule where fragments
split into two equal pieces. In the next section, we show that
stochastic rules result in qualitatively similar behaviors.

\subsection{Random particle splits}

We start with the case where a randomly selected particle splits upon
collision (this is equivalent to having both particles split). For
simplicity, we focus on monodisperse initial conditions where all
particles have unit mass, $m=1$.  Then, a fragment produced by $n$
collision events has mass $m=2^{-n}$.  Let $c_n(t)$ be the density of
such fragments at time $t$. This density evolves according to
\begin{equation}
\label{cn-A-eq}
\frac{d}{dt}\,c_n(t)=N(t)\left[\,2c_{n-1}(t)-c_n(t)\,\right],
\end{equation}
with the total density $N(t)=\sum_{j=0}^{\infty} c_j(t)$. Summing up
Eqs.~(\ref{cn-A-eq}) we indeed recover Eq.~(\ref{N-eq}). Also, the
total mass, $M(t)=\sum_{j=0}^{\infty} 2^{-j}\,c_j(t)$, is conserved,
$M(t)=1$.

In terms of the collision counter, the process is linear,
$\frac{d}{d\tau}\,c_n=2c_{n-1}-c_n$, and subject to the monodisperse
initial conditions $c_n(0)=\delta_{n,0}$, the exact solution is the
Poissonian density \cite{sid}
\begin{equation}
\label{cn-A-sol}
c_n(\tau)=e^{-\tau}\,\frac{(2\tau)^n}{n!}\,.
\end{equation}
At the shattering time $t_c=1$ (corresponding to $\tau=\infty$), the
densities vanish: $c_n(t=1)=0$ for all $n$. Therefore, the fragment
mass density undergoes a first-order (discontinuous) transition,
$M(t)=\Theta(t_c-t)$ with $\Theta$ the Heaviside step function. In
other words, the entire mass is shattered into dust and there are no
particles with positive mass \cite{F,ziff,sid}.

Near the shattering transition, {\it i.e.}, as $\tau\to\infty$, the
mass distribution approaches
\begin{equation}
\label{cn-A-lim}
c_n(\tau)\to \frac{N}{\sqrt{v\tau}}\,\,
G\left(\frac{n-v\tau}{\sqrt{v\tau}}\right)\,,
\end{equation}
where $v=2$ and $G(x)=(2\pi)^{-1/2}\,\exp\left(-x^2/2\right)$ is the
Gaussian distribution.  Since $n=\log_2(1/m)$, the mass distribution
becomes log-normal, a behavior typical to fragmentation and cascade
processes \cite{shr,F,ziff,sid}.

\subsection{Larger particle splits}

Now in a collision the larger particle splits into two equal
pieces. If the colliding particles have the same mass, a randomly
chosen particle splits. The fragment mass density, $c_n(t)$, satisfies
the rate equation
\begin{equation}
\label{cn-B-eq}
\frac{d}{dt}\,c_n=4c_{n-1}A_n-2c_nA_{n+1}+2c_{n-1}^2-c_n^2\,,
\end{equation}
where $A_n$ is the cumulative density of fragments of mass $2^{-n}$
and smaller, $A_n(t)=\sum_{j=n}^\infty c_j(t)$. The initial conditions
are $c_n(0)=\delta_{n,0}$. One can verify that the mass is conserved,
$M(t)=1$, and that the total density is given by Eq.~(\ref{N-sol}).

\begin{figure}
\centerline{\epsfxsize=8.5cm\epsfbox{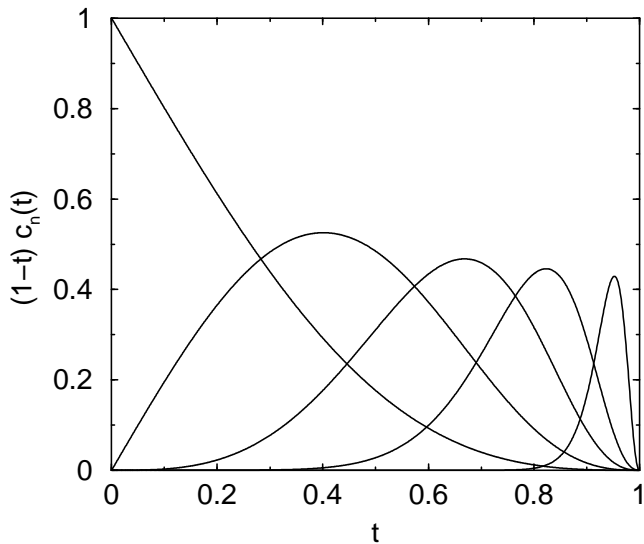}} \caption{The
normalized fragment size distribution. Shown is $N^{-1}c_n(t)$
versus $t$ for $n=0$, $1$, $2$, $4$, and $6$.}
\label{cn} 
\end{figure}

The density $c_0(t)$ of unit mass particles satisfies the Bernoulli equation,
$\frac{d}{dt}\,c_0=c_0^2-2c_0N$. Using Eq.~(\ref{N-sol}) and the initial
condition $c_0(0)=1$ gives
\begin{equation}
\label{c0-B}
c_0(t)=\frac{3 (1-t)^2}{2+(1-t)^3}\,.
\end{equation}
For sufficiently small $n$, one can obtain the leading asymptotic
behavior near the shattering transition.  Since $A_n\to N$ as $t\to 1$
and the last two terms on the right-hand side of Eq.~(\ref{cn-B-eq})
are asymptotically negligible, the rate equations simplify to
$\frac{d}{dt}\,c_n=2N(2c_{n-1}-c_n)$ which are identical (up to the
factor 2) to Eqs.~(\ref{cn-A-eq}). Therefore,
\begin{equation}
\label{cn-B-sol}
c_n(\tau)\propto e^{-2\tau}\,\frac{(4\tau)^n}{n!}.
\end{equation}
Apart from logarithmic corrections, the densities vanish
quadratically: $c_n(t)\propto (1-t)^2$. We conclude that the
shattering transition remains discontinuous (see Fig.~\ref{cn}).
Figure \ref{cn} suggests studying the normalized distribution
$N^{-1}c_n(t)$.  Below, we show that as a function of $\tau$, the
normalized fragment mass distribution follows a universal behavior in
the large-$n$ limit.

\begin{figure}
\centerline{\epsfxsize=8.5cm\epsfbox{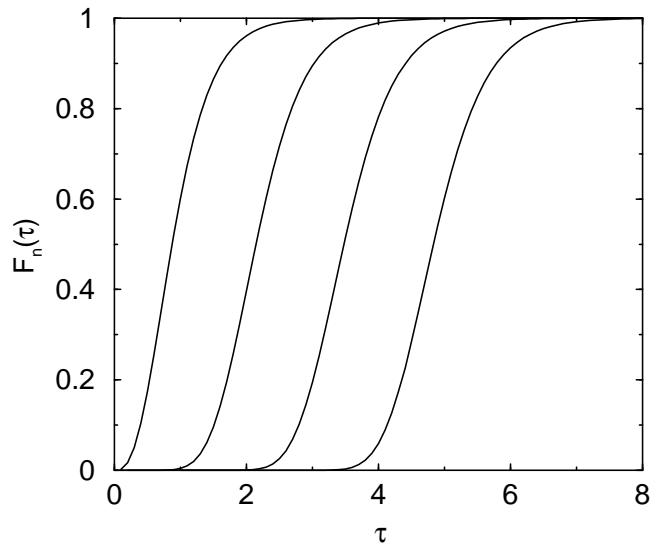}}
\caption{The travelling wave. Shown are numerical solutions of
Eq.~(\ref{Fn-B-eq}) for $n=2$, $4$, $6$, and $8$.}
\label{wave} 
\end{figure}

The rate equations (\ref{cn-B-eq}) simplify in terms of the cumulative
densities:
\begin{equation}
\label{An-B-eq}
\frac{d}{dt}\,A_n=2A_{n-1}^2-A_n^2\,.
\end{equation}
This equation holds for $A_0=N$ if we set $A_{-1}\equiv A_0$.  The
initial conditions are $A_n(0)=\delta_{n,0}$. We characterize time by
the collision counter (\ref{tau}) and normalize the size density by the
total number density, $F_n(\tau)=N^{-1}A_n(t)$. These transformations yield
\begin{equation}
\label{Fn-B-eq}
\frac{d}{d\tau}\,F_n=2F_{n-1}^2-F_n^2-F_n\,.
\end{equation}
Asymptotically, this equation admits a travelling wave solution
$F_n(\tau)\to f(n-v\tau)$ as shown in Fig.~\ref{wave}.  The wave form
$f(x)$ satisfies the difference-differential equation
\begin{equation}
\label{fx-B-eq}
v\frac{d}{d x}f(x)=f(x)+f^2(x)-2f^2(x-1)\,,
\end{equation}
and is subject to the boundary conditions $f(-\infty)=1$ and
$f(\infty)=0$. Remarkably, the velocity $v$ can be determined without
solving the nonlinear and non-local differential equation
(\ref{fx-B-eq}) exactly. It follows from the exponential behavior
attained by $f(x)$ far behind the front: $1-f(x)\sim e^{\lambda x}$ as
$x\to -\infty$. Together with Eq.~(\ref{fx-B-eq}) it yields a
``dispersion'' relation between the velocity $v$ and the decay
coefficient $\lambda$,
\begin{equation}
\label{v-B}
v=\frac{3-4e^{-\lambda}}{\lambda}\,.
\end{equation}
Out of the spectra of possible velocities $v\in (-\infty,v_{\rm
max}]$, the maximal value is selected. At the maximum, we have
$3e^{\lambda}=4(1+\lambda)$, from which $\lambda\cong 0.961279$ and
$v\cong 1.52961$.  Alternatively, the velocity is the smaller root of
$v\ln(4e/v)=3$.

Velocity selection underlies numerous situations, yet it has been
rigorously established only for a few non-linear parabolic partial
differential equations, typically occurring in reaction-diffusion
problems \cite{bar,jdm,mb,vs,bd,evs}.  Recently, velocity selection
has been also applied to a host of difference and
difference-differential equations \cite{bbdk,mk,km,mk1,bkm} including
a linear fragmentation process \cite{km}.  Typically, the selected
velocity gives key physical characteristics such as the growth
velocity of a surface in deposition processes \cite{bbdk} or the
extremal heights of random trees \cite{bkm}.

The typical behavior of the fragment mass density follows from 
the travelling wave form 
\begin{equation}
\label{g} 
c_n\to N\,g(n-v\tau)\,,
\end{equation}
with $g(x)=f(x)-f(x+1)$. The front location $n_*\approx v\tau$
characterizes typical fragments and the typical mass $m_*=2^{-n_*}$
shrinks as 
\begin{equation}
\label{sigma} 
m_*\sim (1-t)^{\sigma}\,
\end{equation}
with $\sigma=v\ln 2\cong 1.06024$ as $t\to 1$.  The typical mass
decays slower than in model A where $\sigma=2\ln2\cong 1.38629$.
Another difference between models A and B is manifested by the width:
In contrast with the diffusive broadening in model A the width
saturates at a finite value in model B.  Yet, fundamentally the
shattering transitions are the same in both models --- the entire
system is instantly transformed into dust at the transition point.

The extremal behavior of the fragment mass density follows from the
tails of $f(x)$. The behavior far ahead of the wave front ($x\to
\infty$) is a sharp double exponential decay, as implied by the
leading terms in Eq.~(\ref{fx-B-eq}), $v\frac{d}{dx}f(x)=-2f^2(x-1)$.  
In summary, the extremal behaviors are 
\begin{equation}
\label{fx-B-lim} 
f(x)\sim \cases
{1-C_1\,e^{\lambda x}&$x\to-\infty,$\cr
 2^x\,\exp(-C_2\, 2^x)&$x\to+\infty.$}
\end{equation}

We now re-express the mass distribution in terms of the ordinary mass
variable $m=2^{-n}$. The two distributions are related via
$c(m)dm=c_ndn$.  Near the shattering transition, the mass distribution
attains the scaling form $c(m)\to \frac{N}{m_*}{\cal
F}\left(\frac{m}{m_*}\right)$. Equation (\ref{fx-B-lim}) leads to the
following extremal behaviors of the scaling function:
\begin{equation}
\label{Fz-B-lim} 
{\cal F}(z)\sim \cases
{z^{-\alpha} &$z\gg 1$,\cr
z^{-2}\exp\left(-C_2\,z^{-1}\right)&$z\ll 1$;}
\end{equation}
with $\alpha=1+\lambda/\ln 2\cong 2.38683$. Hence, large masses
(relative to the typical mass) are suppressed algebraically, while
small masses are suppressed exponentially.

Generally, in fragmentation processes the mass distribution has a
scaling form and this is indeed the case for collision-induced
fragmentation. However, the nonlinear nature of the process results in
qualitative changes to the scaling behavior. The similarity solutions
have two scales characterizing the front location and fluctuations
around it in the linear case (model A).  In contrast, only a single
scale underlies similarity solutions in the nonlinear case (model B).

\subsection{Smaller particle splits}

When the smaller particle splits upon collision the fragment size
densities satisfy the rate equations
\begin{equation}
\label{cn-C-eq}
\frac{d}{dt}c_n=4c_{n-1}B_{n-1}-2c_nB_n+2c_{n-1}^2-c_n^2\,,
\end{equation}
where $B_n=\sum_{j=0}^{n-1} c_j$ is the cumulative density of
particles with mass larger than $2^{-n}$. 

The density of unit mass particles is readily found by  
solving $\dot c_0=-c_0^2$.  The next density can be found as well 
\begin{eqnarray}
\label{c0-C}
c_0(t)&=&\frac{1}{1+t}\,,\\
c_1(t)&=&\frac{2}{1+t}\,\,\frac{(1+t)^3-1}{2(1+t)^3+1}\,\nonumber.
\end{eqnarray}
These explicit results already demonstrate that densities are {\em positive}
at all times. Hence, the total mass density $M(t)$ also remains positive
after the shattering transition.

The kinetics just below and at the shattering transition can be
determined using the travelling wave behavior. The cumulative
distribution obeys \hbox{$\frac{d}{dt}\,B_n=2B_{n-1}^2-B_n^2$} which
is identical to Eq.~(\ref{An-B-eq}); the initial conditions, however,
are different: $B_n(0)=1-\delta_{n,0}$.  The transformed distribution
$F_n(\tau)=N^{-1}B_n$ again evolves according to Eq.~(\ref{Fn-B-eq}).
Asymptotically, it admits a travelling wave solution, $F_n(\tau)\to
f(n-v\tau)$, with the wave form $f(x)$ satisfying Eq.~(\ref{fx-B-eq}).
However, the boundary conditions are reversed, $F(-\infty)=0$ and
$F(\infty)=1$, leading to different quantitative and qualitative
results.

Both extremal behaviors are now exponential
\begin{equation}
\label{fx-C-lim}
f(x)\sim\cases{
e^{x/v}&$x\to -\infty$,\cr 
1-e^{-\lambda x}&$x\to +\infty$.}
\end{equation}
The behavior far ahead of the front is used to determine the
velocity.  The dispersion relation is 
\begin{equation}
\label{v-C}
v=\frac{4e^\lambda-3}{\lambda}\,,
\end{equation}
and the extremum selection principle gives $\lambda\cong 0.58013$ and
$v\cong 7.14509$. Numerically, we confirmed this velocity to within
0.01\%. Interestingly, $v$ is the larger root of the same (as in model
B) equation $v\ln(4e/v)=3$.  We note that the velocities satisfy
$v_B<v_A<v_C$.

\begin{figure}
\centerline{\epsfxsize=8.5cm\epsfbox{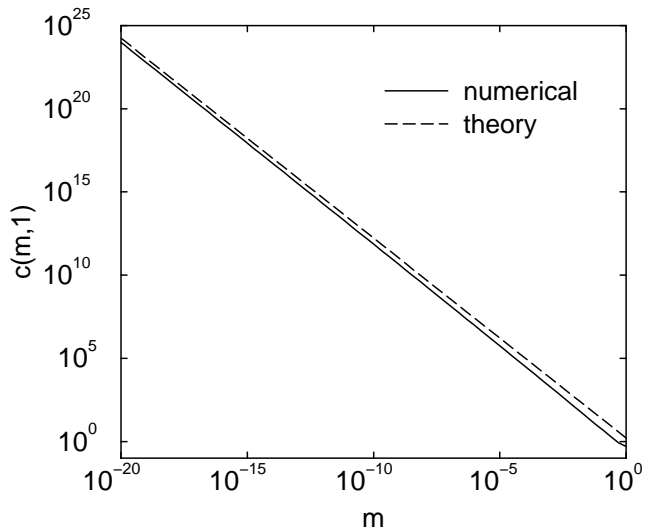}}
\caption{The mass distribution at the shattering time. Numerical
integration of the rate equations (\ref{cn-C-eq}) are compared with
the theoretical prediction (\ref{cm1-C-lim}).}
\label{cm1}
\end{figure}

The fragment size distribution follows the travelling wave form
(\ref{g}) with $g(x)=f(x+1)-f(x)$.  The typical mass shrinks according
to (\ref{sigma}) with $\sigma=v\ln 2\cong 4.9526$ near the shattering
point. The exponential tails of the wave form imply algebraic tails
for the scaling function underlying the mass distribution
\begin{equation}
\label{Fz-C-lim} {\cal F}(z)\sim \cases
{z^{-\alpha} & $z\gg 1$,\cr
z^{-\beta} &$z\ll 1$;}
\end{equation}
with $\alpha=1+(v\ln 2)^{-1}\cong 1.20191$ and $\beta=1-\lambda/\ln 2\cong 0.163049$. 

Our major result is that the mass distribution diverges algebraically at 
the transition time \cite{note1}:  
\begin{equation}
\label{cm1-C-lim}
c(m,1)\sim m^{-\alpha},
\end{equation}
for $m\to 0$ with the transcendental exponent $\alpha=1.20191$
(Fig.~3).  This behavior can be obtained from the large-$z$ behavior
of ${\cal F}(z)$.  Although in general the travelling wave form
implies time-dependent densities, when $z\to\infty$, the mass
densities become {\it stationary}.

\begin{figure}
\centerline{\epsfxsize=8.5cm\epsfbox{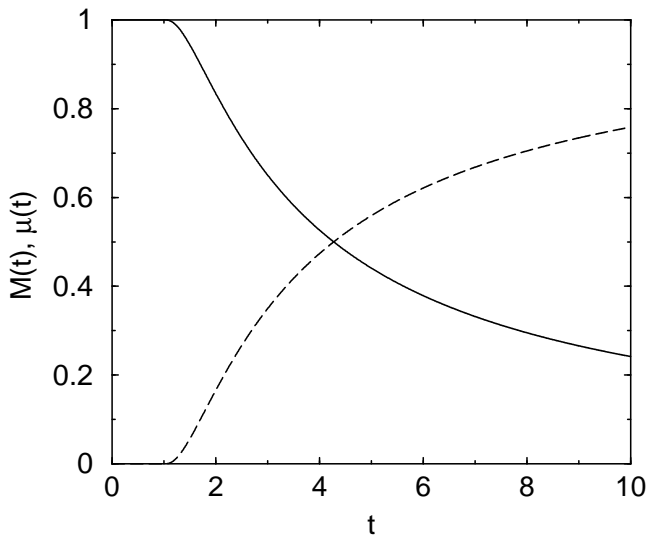}}
\caption{Fragment versus dust mass. Shown are the fragment mass $M(t)$
(solid line) and the dust mass $\mu(t)$ (dashed line) versus time $t$.}
\label{masses}
\end{figure}

Model C exhibits a rich post-transition behavior.  The explicit
solutions (\ref{c0-C}) suggest that $c_n\simeq \gamma_n\,t^{-1}$ when
$t\to\infty$. Indeed, this behavior is compatible with
Eqs.~(\ref{cn-C-eq}) and the cumulative amplitudes
$\Gamma_n=\sum_{j=0}^n\gamma_j$ satisfy the recursion relation
$\Gamma_n^2-\Gamma_n=2\Gamma_{n-1}^2$ with $\Gamma_0=1$. The
amplitudes grow exponentially, $\gamma_n\sim \Gamma_n\sim 2^{n/2}$.
Summing over densities, the total fragment mass decays as 
\begin{equation}
\label{long}
M(t)\simeq C\,t^{-1}\quad {\rm as}\quad t\to \infty,
\end{equation}
with $C=\sum_{n=0}^\infty 2^{-n}\gamma_n\cong 2.66084$.  Thus, the
total fragment mass remains positive at all times. The dust mass,
$\mu(t)=1-M(t)$, vanishes at the shattering time, $\mu(1)=0$, and it
gradually increases for $t>1$ (Fig.~\ref{masses}).  Only in the long time
limit it accounts for the entire mass in the system.  We conclude that
in model C, the shattering transition is continuous.

\begin{figure}
\centerline{\epsfxsize=8.5cm\epsfbox{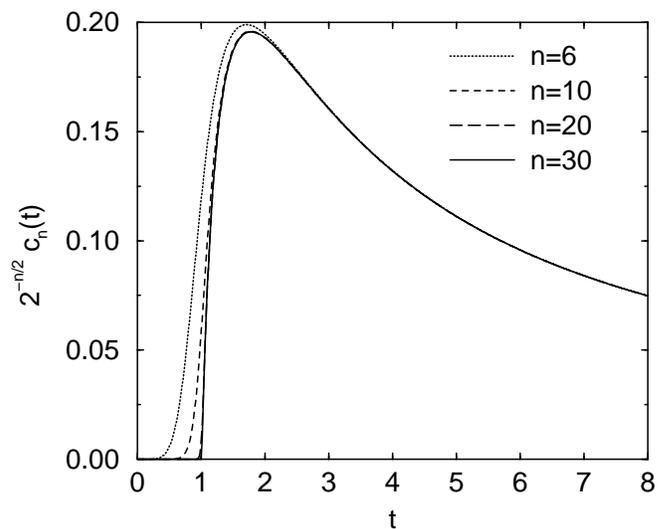}}
\caption{The asymptotic behavior of the size density. 
Shown is $2^{-n/2}c_n(t)$ versus $t$ for $n=6$, $10$, $20$, and $30$.}
\label{u}
\end{figure}

Numerically, we observe that for sufficiently large $n$, the densities 
follow a universal behavior (Fig.~\ref{u})
\begin{equation}
\label{cn-C-lim}
c_n(t)\to 2^{n/2}\,u(t).
\end{equation}
While this ansatz is asymptotic with respect to $n$, it holds for {\em
all} times. The function $u(t)$ vanishes below the shattering time and
grows linearly afterwords, $u(t)\sim (t-1)$ for $t-1\to 0$.  Hence,
this function plays the role of an order parameter.  Note also that
$u(t)\sim t^{-1}$ as $t\to\infty$.

The order parameter and the total dust mass are intimately related.
Consider the total mass density of fragments of mass $2^{-k}$ or
larger: $M^{(k)}(t)=\sum_{n=0}^k 2^{-n}\,c_n(t)$. From the rate
equations (\ref{cn-C-eq}), this mass density decreases according to
\begin{equation}
\label{Mk}
\frac{d}{dt}\,M^{(k)}(t)=-2^{-k}c_k\left(2B_k+c_k\right).
\end{equation}
The flux of mass from fragments into dust is simply
\hbox{$\frac{d}{dt}\,\mu=-\lim_{k\to\infty} \frac{d}{dt}\,M^{(k)}$}.
Using Eq.~(\ref{cn-C-lim}), the right-hand side of Eq.~(\ref{Mk})
approaches $\left(3+2\sqrt{2}\,\right) u^2(t)$ in the limit
$k\to\infty$ and therefore, 
\begin{equation}
\label{mu}
\frac{d}{dt}\,\mu(t)=\big(3+2\sqrt{2}\big) u^2(t)\,.
\end{equation}
This in turn shows that the dust mass grows according to $\mu(t)\sim
(t-1)^3$ past the transition.

\section{Stochastic Fragmentation}

We now briefly describe a generalized collision-induced fragmentation
process where splitting is stochastic. Specifically, a particle of
mass $m$ splits into two fragments of mass $m'$ and $m-m'$ with $m'$
chosen stochastically from the interval $0<m'<m$ according to some
fixed distribution. We focus on the simplest case of uniform
splitting, i.e., $m'$ is chosen uniformly in $[0,m]$.

\subsection{Model A}

When a randomly selected particle splits, the mass density $c(m,\tau)$
satisfies 
\begin{equation}
\label{cm-AA-eq}
\frac{\partial}{\partial\tau}\,c(m,\tau)=-c(m,\tau)+
2\int_m^\infty \frac{dm'}{m'}\,\,c(m',\tau)\,.
\end{equation}
The kernel $1/m'$ reflects the uniform splitting probability and the
collision rate $N$ is absorbed by the collision counter $\tau$.  This
equation is solved using the Mellin transform and for the monodisperse
initial condition, $c(m,0)=\delta(m-1)$, one finds 
\begin{eqnarray*}
c(m,\tau)=e^{-\tau}\delta(m\!-\!1)+
e^{-\tau}\sqrt{\frac{2\tau}{\ln\frac{1}{m}}}\,\,
I_1\left[\,\sqrt{8\tau \ln\frac{1}{m}}\,\right]
\end{eqnarray*}
with $I_1$ the modified Bessel function. The first term on the
right-hand side simply describes the density of particles that have
yet to collide.  The second term simplifies asymptotically. Making the
transformation $m=e^{-n}$ leads to a normal distribution as in
(\ref{cn-A-lim}) with the propagation velocity $v=\frac{32}{9}$.
Thus, the qualitative behavior is independent of the splitting rule.

\subsection{Model B}

When the larger of the two fragments splits, the rate equations for 
the mass density are 
\begin{equation}
\label{cm-BB-eq}
\frac{\partial}{\partial t}c(m)
=4\int_m^\infty \frac{dm'}{m'} c(m')A(m')-2c(m)A(m)\,,
\end{equation}
with the cumulative density $A(m)=\int_0^m dm' c(m')$. We employ the
same transformations used in the deterministic case.  Characterizing
the mass $m$ by ``index'' $n$ via $m=e^{-n}$, the fragment size
density, $c(n)$, evolves according to
\begin{equation}
\label{cn-BB-eq}
\frac{\partial}{\partial t}\,c(n)
=4\int_0^n\,dn' e^{n'-n}c(n')A(n')-2c(n)A(n)
\end{equation}
with $A(n)=\int_n^{\infty} dn' c(n')$. This cumulative distribution
satisfies 
\begin{equation}
\label{An-BB-eq}
\frac{\partial}{\partial t}\,A(n)=
A^2(n)-2\int_0^n dn'\,e^{n'-n}\frac{\partial}{\partial n'}A^2(n')\,.
\end{equation}
Expressing time in units of the collision counter and normalizing by the
total density, $F(n,\tau)=N^{-1}A(n)$, we transform Eq.~(\ref{An-BB-eq}) into
\begin{eqnarray*}
\frac{\partial}{\partial\tau}\,F(n)=
F^2(n)-F(n)-2\int_0^ndn'\,e^{n'-n}\frac{\partial}{\partial n'}F^2(n')\,.
\end{eqnarray*}
Seeking a travelling wave solution $F(n,\tau)\to f(n-v\tau)$ yields
the non-linear integro-differential equation
\begin{equation}
\label{fx-BB-eq}
v{d\over dx}f(x)=f(x)-f^2(x)+2\int_{-\infty}^x dy\, 
e^{y-x}\frac{d}{dy}f^2(y)\,,
\end{equation}
subject to the boundary conditions $f(-\infty)=0$ and $f(\infty)=1$.  The
exponential decay $1-f(x)\sim \exp(\lambda x)$ as $x\to \infty$ gives the
dispersion relation $v=4(1+\lambda)^{-1}-\lambda^{-1}$ and the extremum
selection principle yields $\lambda=v=1$. Close to the shattering transition,
the typical mass is proportional to the average mass, $m_*\sim (1-t)$
\cite{note}.  The mass densities behave as in the deterministic case and the
extremal behaviors (\ref{Fz-B-lim}) are recovered with $\alpha=2$.  The
nature of the transition is discontinuous, as in the deterministic case.

\subsection{Model C}

When the smaller particle splits upon collision, the rate equations
for the mass density are
\begin{equation}
\label{cm-CC-eq}
\frac{\partial }{\partial t}c(m)
=4\int_m^\infty \frac{dm'}{m'}\,\,c(m')B(m')-2c(m')B(m')
\end{equation}
with $B(m,t)=\int_m^\infty dm'\, c(m',t)$.  In terms of the index $n$, the
cumulative density $B(n,t)=\int_0^n dn' c(n')$ obeys Eq.~(\ref{An-BB-eq}).
The normalized cumulative density again admits the travelling wave form.  The
velocity and decay rate are $v=9$ and $\lambda=\frac{1}{3}$. At the
shattering time, the (finite) mass distribution diverges algebraically:
$c(m,1)\sim m^{-\alpha}$ with $\alpha=10/9$.  Past the shattering transition,
the asymptotic ansatz $c(m,t)\to m^{-3/2}\,u(t)$ holds for small masses and
the dust mass is related to the order parameter via $\frac{d}{dt}\,\mu=2u^2$.
We conclude that qualitatively, the shattering transition is similar to the
deterministic case.

\section{Discussion}

We investigated kinetic properties of collision-induced fragmentation
processes.  Generally, the mass is transferred from finite fragments into
infinitesimal dust in a finite time. The nature of the shattering transition
depends on the fragmentation process. When the larger of the colliding
particles splits or when a randomly selected one splits, the transition is
discontinuous and the entire mass is transformed into dust instantaneously.
When the smaller particle splits, the transition is continuous, with the dust
accumulating gradually past the shattering transition. In this case, finite
fragments always carry a non-zero fraction of the mass.

Model A is essentially linear and thus solvable.  For models B and C
the nonlinear and non-local governing equations can not be solved in a
closed form. Nevertheless, in the vicinity of the shattering
transition we were able to obtain the most important characteristics
analytically by utilizing the traveling wave form of the fragment mass
density. The mass distribution follows a scaling behavior with a
single characteristic scale, in contrast with the two scales found for
linear processes.

For model C, the post-shattering behavior is nontrivial.  At the
transition point, the mass distribution decays algebraically,
$c(m)\sim m^{-\alpha}$, with a transcendental exponent
$\alpha=1.20191\ldots$ in deterministic fragmentation and a rational
exponent $\alpha=10/9$ in stochastic fragmentation.  We have also
demonstrated that the mass densities exhibit universal asymptotic
behavior (\ref{cn-C-lim}) in the post-shattering region.

A challenging open problem is the complete post-shattering behavior in
model C, for example, the time dependent dust mass. This is largely a
mathematical problem since physically, the breakage of sufficiently
small fragments is impossible. For instance, micro-cracks on the
surface of the fragment are often precursors for breakage.  The number
of such surface defects is proportional to the surface area, so
sufficiently small fragments are effectively unbreakable.

We focused on the leading asymptotic behavior. There are however
corrections to the linear front propagation \cite{mb,vs,bd,evs}.  The
traveling wave solution is actually a function of $x=n-X(\tau)$ with
the position of the front $X(\tau)$ given by
\hbox{$X(\tau)=v\tau\pm{3\over 2\lambda}\,\ln \tau+{\cal O}(1)$}.  The
plus and minus signs correspond to model B and C, respectively. This
translates to a logarithmic correction to the typical mass
(\ref{sigma}).

We treated the problem using a mean-field rate equation approach.
Thus we ignored correlations between the colliding particles. In
principle, spatial correlations may be important up to some critical
dimension beyond which they can indeed be ignored.  The analysis of
this possibility requires a more complete description of the
process. Particularly, one must specify the transport mechanism.

Collision-induced fragmentation arises most naturally in processes
where particles moves ballistically between collisions. Using
dimensional analysis we argue that the shattering transition always
occur in ballistic fragmentation. The typical mean free time $T$,
velocity $v$, particle cross section $s$, and number density $N$ are
related via $NvTs\sim 1$. Mass conservation implies $m\sim N^{-1}$
(here $m$ is the typical mass), while energy conservation gives
$v\sim 1$.  Finally $m\sim s^{d/(d-1)}$ yields $s\sim N^{-1+1/d}$.  In
particular, $T\sim N^{-1/d}$.  The particle density evolves according
to $\frac{d}{dt}\,N=N/T$, or $\frac{d}{d t}\,N\sim N^{1+1/d}$ from
which $N\sim (t_c-t)^{-d}$.  This suggests that in ballistic
fragmentation, the shattering transition occurs in arbitrary dimension
$d$.  Using effective $d$-dimensional collision rates ($\propto
N^{1/d}$) one can convert the ``one-dimensional'' results in this
study into a general mean-field theory.

\acknowledgments
We are grateful to K.~Kornev for very fruitful remarks. 
This research was supported by DOE (W-7405-ENG-36).

\end{document}